\documentclass[
  aps,
  prd,
  preprint,
  titlepage,
  superscriptaddress,
  amsmath,
  amssymb
]{revtex4-2}

\usepackage[T1]{fontenc}
\usepackage[utf8]{inputenc}
\usepackage{bm}
\usepackage{slashed}
\usepackage{graphicx}
\usepackage[hidelinks]{hyperref}

\hypersetup{
  pdftitle={Quantifying TMD factorization breaking at tree level},
  pdfauthor={Jian Zhou}
}

\begin{document}

\title{Quantifying TMD factorization breaking at tree level}

\author{Jian Zhou}
\email{jzhou@sdu.edu.cn}
\affiliation{Key Laboratory of Particle Physics and Particle Irradiation (MOE),
Institute of Frontier and Interdisciplinary Science, Shandong University,
Qingdao, Shandong 266237, China}
\affiliation{Southern Center for Nuclear-Science Theory (SCNT), Institute of
Modern Physics, Chinese Academy of Sciences, Huizhou, Guangdong 516000, China}

\begin{abstract}
Rogers and Mulders identified color entanglement as a mechanism that
breaks transverse-momentum-dependent (TMD) factorization in nearly
back-to-back dijet production. Using the double-Sivers asymmetry in
their model, we quantify the contribution of color entanglement in the
intermediate region $\Lambda_{\rm QCD}\ll q_T\ll P_T$, where $q_T$ is
the transverse-momentum imbalance of the dijet and $P_T$ is the hard
transverse momentum. We compare the collinear twist-three result with
the TMD-factorization prediction, expressing both in terms of the Qiu--Sterman functions. In the absence of color entanglement, the
factorized prediction would involve the Sivers functions of
semi-inclusive deep-inelastic scattering (SIDIS). Diagrams without crossed attachments reproduce the
hard coefficient obtained from the perturbative tails of these Sivers
functions. Crossed
diagrams, in which each coherent gluon attaches to the other hadron's
quark line, contribute an extra term equal to $4/(N_c^2-1)$ times this
coefficient. 
\end{abstract}

\maketitle

\section{Introduction}
\label{sec:introduction}

Transverse-momentum-dependent (TMD) parton distributions are central to
the three-dimensional imaging of the proton. The gauge links
in the TMD operator definitions encode initial- and final-state
interactions, and their path structure has observable consequences.
One such consequence is the predicted sign change of T-odd TMDs between
semi-inclusive deep-inelastic scattering (SIDIS) and Drell--Yan
production, which arises because the gauge link points to the future in
SIDIS and to the past in Drell--Yan \cite{Collins:2002kn,Brodsky:2002rv}.

The situation is more complicated for back-to-back jet
production in hadron--hadron collisions, where gluons from either
hadron can attach to several incoming and outgoing parton lines.
The resulting process-dependent gauge links and gluonic-pole factors
were derived in
Refs.~\cite{Bomhof:2004aw,Bomhof:2006dp,Bomhof:2006ra}.
These results imply that conventional TMD factorization breaks down
\cite{Collins:2007nk,Collins:2007jp}. This prompted attempts to preserve
predictive power through a generalized TMD factorization with
process-dependent gauge links.
Rogers and Mulders \cite{Rogers:2010dm} showed, however, that
simultaneous gluon attachments from the two hadrons can violate even
this generalized form of factorization. In their double-Sivers example,
the crossed diagrams correlate the colors of the two hadrons, and the
corresponding pole contributions do not cancel in the sum over all cuts.
This correlation is known as color entanglement. Factorization breaking
can also give rise to new spin asymmetries at leading power
\cite{Rogers:2013zha}. Color entanglement has been further studied in
Refs.~\cite{Schafer:2014zea,Schafer:2014xpa,Zhou:2015ima,
Zhou:2017sdx,Zhou:2017mpw}, and its cancellation in Drell--Yan
production was discussed in Ref.~\cite{Boer:2017hqr}.

In this work, we take the double-Sivers asymmetry in the Rogers--Mulders
model as an example and quantify the factorization breaking caused by
color entanglement. We focus on the region
$\Lambda_{\rm QCD}\ll q_T\ll P_T$, where $q_T$ is the dijet imbalance
and $P_T$ is the hard transverse momentum. This intermediate region permits
a perturbative calculation, while TMD factorization would be expected to
apply \cite{Ji:2006ub,Qiu:2007ar,Qiu:2007ey}.
We compute this asymmetry in the collinear twist-three approach
\cite{Zhou:2010ui,Schafer:2013wca,Ma:2014uma,Chen:2016dnp,Chen:2017lvx,
Rein:2025pwu,Rein:2025qhe} and compare it with the
TMD-factorization result built from the perturbative tails of the SIDIS
Sivers functions.
Both calculations are expressed in terms of the same Qiu--Sterman
functions, so any difference between them resides entirely in the hard
coefficients.

To isolate the color-entanglement contribution, we use a model in which
the hard scattering proceeds through photon exchange. The perturbative imbalance is generated by real-photon
radiation, which carries no color and therefore leaves the color flow
of the hard scattering unchanged.
We find that contributions without crossed attachments reproduce the
hard coefficient obtained from SIDIS Sivers functions, whereas the
crossed attachments contribute an extra term equal to $4/(N_c^2-1)$
times that coefficient. At leading power, the collinear
result is therefore enhanced relative to the TMD-factorization result
by the factor
\begin{equation}
 R=1+\frac{4}{N_c^2-1}.
 \label{eq:ratio}
\end{equation}

\section{Setup and the TMD-factorization prediction}
\label{sec:framework}
\label{sec:results}

We adopt the Rogers--Mulders model \cite{Rogers:2010dm} and include the
emission of a real photon, which generates a perturbatively large dijet
imbalance,
$\Lambda_{\rm QCD}\ll q_T\ll P_T$. The partonic process is
\begin{equation}
 q_f(x_AP_A)+q_{f'}(x_BP_B)
 \longrightarrow q_f(p_1)+q_{f'}(p_2)+\gamma(k),
 \qquad f\ne f',
 \label{eq:process}
\end{equation}
where the hard scattering proceeds through virtual-photon exchange
between the two quark lines. For simplicity, we take $e_f^2=e_{f'}^2$.
The incoming hadrons have momenta $P_A$ and $P_B$ and transverse spins
$S_A$ and $S_B$.
We label each jet by the flavor of the quark that initiates it. We use the light-cone components
$v^\pm=v^0\pm v^3$ and the metric
$(+,-,-,-)$. The dijet imbalance
and the hard transverse momentum are defined by
\begin{equation}
 \bm q_T=\bm p_{1T}+\bm p_{2T},\qquad
 \bm P_T=\frac{\bm p_{1T}-\bm p_{2T}}{2},\qquad
 \Lambda_{\rm QCD}\ll q_T\ll P_T.
 \label{eq:measurement}
\end{equation}

The double-spin cross section is defined by
\begin{equation}
 d\Delta\Sigma
 =\frac14\big[d\Sigma(S_A,S_B)-d\Sigma(-S_A,S_B)
 -d\Sigma(S_A,-S_B)+d\Sigma(-S_A,-S_B)\big].
 \label{eq:double-spin}
\end{equation}
We compute the leading-power double-Sivers term, which scales as
$1/q_T^4$ up to logarithms. The collinear result $\Delta\Sigma$ includes
coherent-gluon attachments to both the incoming and the outgoing quark
lines.
We denote by $\Delta\Sigma^{\mathrm{ref}}$ the corresponding
TMD-factorization result built from SIDIS Sivers functions.

For each beam, let $n$ be a lightlike vector with $P\cdot n=1$.
The dual transverse spin vectors are defined by
$\tilde S_A^i\equiv\epsilon_A^{ij}S_{A j}$ and
$\tilde S_B^i\equiv\epsilon_B^{ij}S_{B j}$, where
$\epsilon_A^{ij}=\epsilon^{ij\rho\sigma}P_{A\rho}n_{A\sigma}$ and
$\epsilon_B^{ij}=\epsilon^{ij\rho\sigma}P_{B\rho}n_{B\sigma}$.
When the beam label is omitted, we write
$\tilde S_T^i\equiv\epsilon_T^{ij}S_{Tj}$.

We consider the leading-power contribution from photon radiation
collinear to either incoming quark, characterized by $k^+$ for emission
along beam $A$ and by $k^-$ for emission along beam $B$. At zero incoming transverse momentum, the photon on-shell
condition, the momentum fractions carried by the quarks entering the
Born subprocess, and the partonic center-of-mass energy squared are
\begin{equation}
 k^+k^-=q_T^2,\qquad
 z_A=1-\frac{k^+}{x_AP_A^+},\qquad
 z_B=1-\frac{k^-}{x_BP_B^-},\qquad
 \hat{s}=x_Ax_BP_A^+P_B^-.
 \label{eq:longitudinal-kinematics}
\end{equation}
Keeping the cross section differential in the jet rapidities $y_1,y_2$,
we obtain the quark momentum fractions fixed by the jet kinematics
at leading power,
\begin{equation}
 \xi_A=\frac{P_T(e^{y_1}+e^{y_2})}{P_A^+},\qquad
 \xi_B=\frac{P_T(e^{-y_1}+e^{-y_2})}{P_B^-}.
 \label{eq:measured-fractions}
\end{equation}
Thus $z_A=\xi_A/x_A$ and $z_B=\xi_B/x_B$.
The expansion is understood in the distributional sense and excludes
the equal-rapidity limit, where the longitudinal Jacobian vanishes.
The normalization is fixed by the Born cross section,
\begin{equation}
 \frac{d\Sigma_{\rm Born}}{dy_1\,dy_2\,d^2\bm P_T}
 =\sigma_0\,\xi_A f_f(\xi_A)\,\xi_B f_{f'}(\xi_B),
 \label{eq:born-coefficient}
\end{equation}
where $\sigma_0=\frac{2\alpha_{\mathrm{em}}^2e_f^4}{P_T^4}
 \left[\frac{1}{(1+e^{y_2-y_1})^2}
       +\frac{1}{(1+e^{y_2-y_1})^4}\right]$.
The double-Sivers contribution to the reference cross section is given
by the convolution
\begin{align}
 &\frac{d\Delta\Sigma^{\mathrm{ref}}}
 {dy_1\,dy_2\,d^2\bm P_T\,d^2\bm q_T}
 =\sigma_0\xi_A\xi_B\int d^2\bm k_{AT}\,d^2\bm k_{BT}\,
 \delta^{(2)}(\bm q_T-\bm k_{AT}-\bm k_{BT})\nonumber\\
 &\qquad\times
 \left[-\frac{\bm k_{AT}\cdot\tilde{\bm S}_A}{M_A}
 f_{1T,A}^{\perp}(\xi_A,k_{AT}^2)\right]
 \left[-\frac{\bm k_{BT}\cdot\tilde{\bm S}_B}{M_B}
 f_{1T,B}^{\perp}(\xi_B,k_{BT}^2)\right].
 \label{eq:reference-convolution}
\end{align}
Here $f_{1T,A}^{\perp}$ and $f_{1T,B}^{\perp}$ are the soft-subtracted,
ultraviolet-renormalized SIDIS Sivers functions.

For color-singlet exchange, the additional Wilson loops enter as color
traces, whose one-gluon terms vanish because $\operatorname{Tr}t^a=0$
\cite{Rogers:2010dm}. At this order, each hadron's Sivers function
therefore has the same future-pointing gauge link as in SIDIS. At
perturbative $q_T$, the double-Sivers contribution is obtained by
combining the photon-induced perturbative tail of one hadron's Sivers
function with the tree-level transverse moment of the other
spin-dependent factor in Eq.~\eqref{eq:reference-convolution}.
Operator matching gives this moment as $\tilde S_T^iT_F(x,x)/2$
\cite{Boer:2003cm}, with $T_F$ defined in
Eq.~\eqref{eq:TF-definition}.

We use the $\delta$ regulator and soft-subtraction prescription of
Ref.~\cite{Scimemi:2019gge}. In impact-parameter space, the unsubtracted
correlator is divided by the square root of the vacuum soft factor and
then UV renormalized.
Equation~\eqref{eq:reference-convolution} therefore contains no separate
soft factor. We set the renormalization and rapidity scales to
$\mu^2=\zeta=\hat{s}$.

Following the matching procedure used in QCD, we compute within the
present model the Sivers tail generated by collinear photon radiation \cite{Ji:2006vf,Ji:2006br,Scimemi:2019gge}.
The photon-induced Sivers tail is
\begin{equation}
 f_{1T}^{\perp}(\xi,k_T^2)
 =-\frac{\alpha_{\mathrm{em}}e_f^2M}{2\pi^2k_T^4}
 \int_{\xi}^1\frac{dx}{x}\,T_F(x,x)
 \bigg\{\frac{1+z^2}{(1-z)_+}+\delta(1-z)
 \left[\ln\frac{\hat{s}}{k_T^2}+1\right]\bigg\}.
 \label{eq:photon-tail}
\end{equation}
Here $z=\xi/x$.
The Sivers tail contains no off-diagonal hard-pole term and
depends only on $T_F(x,x)$.
In QCD this hard-pole term has a purely non-Abelian color factor and is
therefore absent for photon radiation.

Substituting Eq.~\eqref{eq:photon-tail} into
Eq.~\eqref{eq:reference-convolution} and adding the contribution with
the beams interchanged gives
\begin{align}
 &\frac{d\Delta\Sigma^{\mathrm{ref}}}
 {dy_1\,dy_2\,d^2\bm P_T\,d^2\bm q_T}
 =\sigma_0\frac{\alpha_{\mathrm{em}}e_f^2\xi_A\xi_B}{4\pi^2q_T^4}
 \int_{\xi_A}^1\frac{dx_A}{x_A}
 \int_{\xi_B}^1\frac{dx_B}{x_B}\,
 T_F(x_A,x_A)T_F(x_B,x_B)
 \nonumber\\
 &\quad\times\bigg\{
 \left[\frac{4(\bm q_T\cdot\tilde{\bm S}_A)
                 (\bm q_T\cdot\tilde{\bm S}_B)}{q_T^2}
       -\tilde{\bm S}_A\cdot\tilde{\bm S}_B\right]
 \nonumber\\
 &\qquad\times\bigg[
 \frac{1+z_A^2}{(1-z_A)_+}\delta(1-z_B)
 +\frac{1+z_B^2}{(1-z_B)_+}\delta(1-z_A)
 \nonumber\\
 &\qquad\qquad
 +2\delta(1-z_A)\delta(1-z_B)
 \left(\ln\frac{\hat{s}}{q_T^2}+\frac32\right)\bigg]
 +\delta(1-z_A)\delta(1-z_B)\,
 \tilde{\bm S}_A\cdot\tilde{\bm S}_B\bigg\}.
 \label{eq:reference-result}
\end{align}
We now test this prediction against an independent collinear twist-three
calculation.

\section{Collinear twist-three calculation}
\label{sec:entanglement}

In collinear factorization, the double-Sivers contribution involves one
twist-three quark--gluon correlator from each hadron.  We compute the asymmetry directly in terms of the Qiu--Sterman function, the
twist-three quark--gluon correlator defined as \cite{Qiu:1991pp,Qiu:1991wg,Qiu:1998ia,Ji:2006vf}
\begin{equation}
 \tilde S_T^iT_F(x_1,x_2)
 =-\int\frac{d\lambda\,d\eta}{4\pi}\,
 e^{i\lambda x_1+i\eta(x_2-x_1)}
 \langle P,S_T|\bar\psi(0)\slashed n\,
 g_sF^{in}(\eta n)\psi(\lambda n)|P,S_T\rangle .
 \label{eq:TF-definition}
\end{equation}
Here $F^{in}=F^{i\mu}n_\mu$ is the field strength contracted with the
lightlike vector $n$, and straight gauge links along $n$ connect the
fields. Our conventions are $D_\mu=\partial_\mu-ig_sA_\mu$,
$\epsilon_T^{ij}=\epsilon^{ij\rho\sigma}P_\rho n_\sigma$, and
$\epsilon^{0123}=+1$; hadron and flavor labels are suppressed.

We work in light-cone gauge $n\cdot A=0$. This gauge choice simplifies
the field-strength insertion to a total derivative,
$F^{in}(\eta n)=-dA_\perp^i(\eta n)/d\eta$. To fix the remaining gauge
freedom, we impose the retarded boundary condition
$A_\perp(-\infty)=0$ \cite{Zhou:2009jm}, which gives
\begin{equation}
 A_\perp^i(\eta n)
 =-\int_{-\infty}^{\eta}d\eta'\,F^{in}(\eta'n).
 \label{eq:retarded-potential}
\end{equation}
This condition sets the transverse gauge link at past infinity to the
identity.

Combining this integral with the Fourier phases in
Eq.~\eqref{eq:TF-definition} produces the factor
$-i/(x_2-x_1+i0)$. The matching relation for the coherent gluon field
then reads
\begin{equation}
 \int\frac{d\lambda\,d\eta}{4\pi}\,
 e^{i\lambda x_1+i\eta(x_2-x_1)}
 \langle P,S_T|\bar\psi(0)\slashed n\,
 g_sA_\perp^i(\eta n)\psi(\lambda n)|P,S_T\rangle_{\mathrm{ret}}
 =\frac{i\tilde S_T^iT_F(x_1,x_2)}{x_2-x_1+i0}.
 \label{eq:A-matching}
\end{equation}
The advanced condition $A_\perp(+\infty)=0$ would instead give
$A_\perp^i(\eta n)=\int_{\eta}^{\infty}d\eta'\,F^{in}(\eta'n)$, turning
$+i0$ in Eq.~\eqref{eq:A-matching} into $-i0$.

For the transverse-derivative matrix element, only the T-odd part
contributes to the Sivers effect. This part is given by half the
difference between the retarded and advanced matrix elements.
With the retarded prescription the field strength is integrated over
$(-\infty,\lambda)$; with the advanced one, over $(\lambda,\infty)$ and
with the opposite sign. In the difference, the common
covariant-derivative term with a straight gauge link cancels, leaving
only the insertion
$g_s\int_{-\infty}^{\infty}d\eta\,F^{in}(\eta n)$.
Since there is no Fourier phase in $\eta$ for this integral, it selects
$x_2-x_1=0$ and hence the diagonal function $T_F(x,x)$:
\begin{align}
 &\frac12\int\frac{d\lambda}{4\pi}\,e^{ix\lambda}
 \Big[\langle P,S_T|\bar\psi(0)\slashed n\,
 i\partial_\perp^i\psi(\lambda n)|P,S_T\rangle_{\mathrm{ret}}
 \nonumber\\
 &\qquad-\langle P,S_T|\bar\psi(0)\slashed n\,
 i\partial_\perp^i\psi(\lambda n)|P,S_T\rangle_{\mathrm{adv}}\Big]
 =-\frac12\tilde S_T^iT_F(x,x).
 \label{eq:D-matching}
\end{align}
The minus sign corresponds to a past-pointing link; for the
future-pointing link in SIDIS the sign is reversed, giving
$+\tilde S_T^iT_F(x,x)/2$.

Equations~\eqref{eq:A-matching} and \eqref{eq:D-matching} form the core
of the light-cone-gauge method of
Refs.~\cite{Zhou:2009rp,Zhou:2009jm,Zhou:2010ui}, which we follow here.
When the retarded prescription is used throughout, both the
transverse-derivative and the coherent-gluon contributions match
directly
onto the Qiu--Sterman function. This substantially simplifies the
hard-part calculation compared with the conventional covariant-gauge
approach to collinear twist-three factorization.
A different boundary condition would shift contributions between the
derivative and gluon terms but leave their sum unchanged.

To apply these relations systematically, we split the collinear
expansion for each beam into two types of terms
\cite{Zhou:2009rp,Zhou:2009jm}. The $D$ term contains a
transverse-momentum derivative of the hard part multiplying the
two-quark matrix element with the operator
$\bar\psi(0)\slashed n\,i\partial_\perp^i\psi(\lambda n)$. The $T$ term contains
an explicit coherent transverse-gluon insertion,
$\bar\psi(0)\slashed n\,g_sA_\perp^i(\eta n)\psi(\lambda n)$.
These matrix elements are matched using Eqs.~\eqref{eq:D-matching}
and \eqref{eq:A-matching}, respectively.

Choosing one type of term for each beam gives four combinations: $DD$, $TD$, $DT$, and $TT$, where the first letter
refers to beam $A$. The $TT$ contribution requires special attention because it involves
two coherent gluons, one from each hadron. The two nonvanishing
attachment patterns are denoted by $TT_{\rm same}$ and $TT_{\rm cross}$.
In $TT_{\rm same}$, the beam-$A$ gluon attaches to the beam-$A$ quark
line, and the beam-$B$ gluon attaches to the beam-$B$ line.
In $TT_{\rm cross}$, the attachments are exchanged: the beam-$A$ gluon
connects to the beam-$B$ quark line, and vice versa.

These contributions differ in their color structure. For $DD$,
$TD$, $DT$, and $TT_{\rm same}$, the color factors for the two quark
lines are independent, giving $N_c^2C_F^2$ overall. In $TT_{\rm cross}$,
the two gluon color indices link the color traces of the two quark
lines,
$\sum_{a,b}\operatorname{Tr}(t^at^b)\operatorname{Tr}(t^bt^a)
=(N_c^2-1)/4$. This crossed contraction correlates the colors of the
two hadrons; whether this correlation survives depends on the pole
contributions after the sum over cuts.

\begin{figure}[tb]
\centering
\begin{minipage}[b]{0.48\linewidth}
\centering
\includegraphics[width=0.9\linewidth,height=0.8325\linewidth,keepaspectratio]{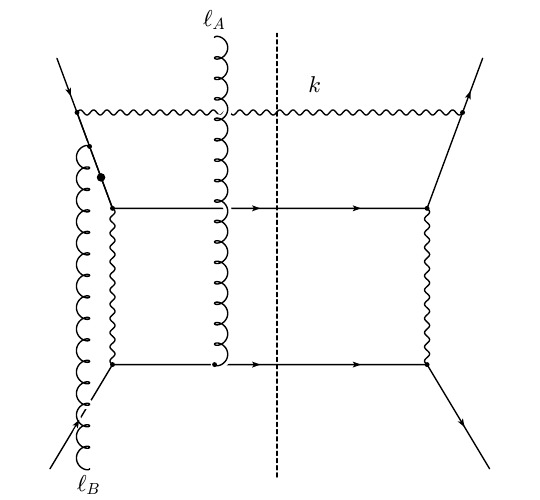}
\par\smallskip(a)
\end{minipage}\hfill
\begin{minipage}[b]{0.48\linewidth}
\centering
\includegraphics[width=0.9\linewidth,height=0.8325\linewidth,keepaspectratio]{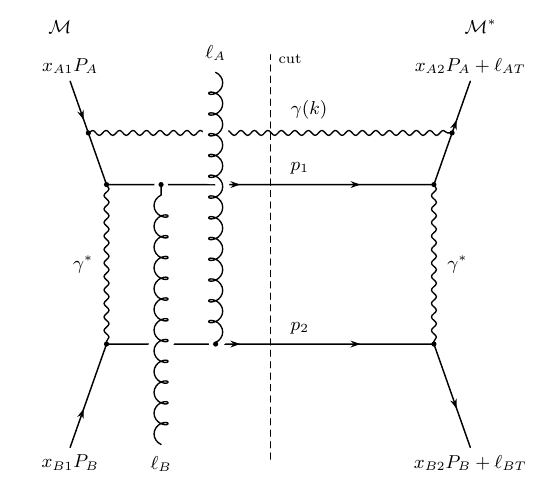}
\par\smallskip(b)
\end{minipage}
\caption{(a) A representative diagram with a hard-gluon pole. The large
solid circle marks the quark propagator carrying
$x_{A1}P_A-k+\ell_B$ between the coherent-gluon insertion and the
hard-scattering vertex.
(b) A representative contribution to $TT_{\rm cross}$. The real photon
is emitted from the incoming beam-$A$ quark on both sides of the cut.
The coherent gluons emerge from the two hadronic correlators and attach
to the opposite outgoing quark lines. Black dots mark vertices; line
crossings without a dot are not vertices.}
\label{fig:crossed-LL}
\label{fig:hard-pole}
\end{figure}

Most of the hard-part calculation follows standard steps, so we discuss
only the extraction of the pole contributions. We first note how the
hard-gluon pole approaches the soft-gluon-pole position at small $q_T$.
In Fig.~\ref{fig:hard-pole}(a), the coherent gluon from beam $B$ attaches
to the beam-$A$ quark line between the hard-scattering vertex and the
photon-emission vertex. For vanishing coherent-gluon transverse
momentum, the marked propagator develops a hard pole at
\begin{equation}
 x_{B2}-x_{B1}
 =\frac{x_{A1}P_A^+k^-}
 {P_B^-(x_{A1}P_A^+-k^+)}.
 \label{eq:hard-pole-position}
\end{equation}
For photon emission collinear to beam $A$, $k^-=q_T^2/k^+$ vanishes as
$q_T\to0$, and the pole approaches $x_{B2}=x_{B1}$. Its
leading-power contribution therefore probes the diagonal Qiu--Sterman
function.

The soft-gluon-pole contribution in the crossed diagram in
Fig.~\ref{fig:crossed-LL}(b) requires more care. Here the photon is emitted from
the incoming beam-$A$ quark on both sides of the cut, while each
coherent gluon attaches to the outgoing quark line from the other
hadron.

The $T$ term does not involve a transverse-momentum derivative.
For a smooth hard part, the coherent-gluon transverse momentum can be
set to zero directly. In the crossed diagram in
Fig.~\ref{fig:crossed-LL}(b), however, the longitudinal integration must
be performed before taking this limit. The reason is
that two longitudinal poles pinch the integration contour as the
coherent-gluon transverse momentum vanishes.

To see these poles explicitly, write the coherent-gluon momenta as
$\ell_A=(x_{A2}-x_{A1})P_A+\ell_{AT}$ and
$\ell_B=(x_{B2}-x_{B1})P_B+\ell_{BT}$. For beam $A$, the gluon
attaches to the outgoing quark with momentum $p_2$, so the adjacent
propagator carries momentum $p_2-\ell_A$. Dropping the beam label on
$x_1,x_2$ and setting $\ell=\ell_A$, $p=p_2$, we combine the pole factor
from Eq.~\eqref{eq:A-matching} with the hard propagator and obtain
\begin{equation}
 \frac{i}
 {(x_2-x_1+i0)[P_A^+p^-(x_2-x_1)-2\bm p_T\cdot\bm\ell_T+\ell_T^2-i0]}.
 \label{eq:final-pole-main}
\end{equation}
The quark--gluon vertex contributes an additional phase factor, which
we keep separate from these poles. For beam $B$, one replaces $P_A^+p^-$
by $P_B^-p^+$ and sets $p=p_1$, $\ell=\ell_B$.

The two longitudinal poles lie on opposite sides of the integration
contour. As we approach the collinear limit, the hard-propagator pole
moves toward the field-strength pole at $x_2=x_1$. Setting $\ell_T=0$
from the beginning would merge these two poles before their pinch
contribution could be evaluated. We therefore keep $\ell_T$ finite,
regulate the two poles with independent positive regulators, integrate
their product, and only then take the collinear limit and expand at
small $q_T$. The two regulators are then taken to zero together, with
their ratio held of order one.

The longitudinal integral contains a soft-gluon-pole (SGP)
contribution. For a function $f(x_2-x_1)$ multiplying the pole factors,
this contribution is
\begin{align}
 &\left[\int d(x_2-x_1)\,
 \frac{i f(x_2-x_1)}
 {(x_2-x_1+i0)[P_A^+p^-(x_2-x_1)-2\bm p_T\cdot\bm\ell_T+\ell_T^2-i0]}
 \right]_{\rm SGP}\nonumber\\
 &\hspace{2cm}
 =-\frac{2\pi f(0)}
 {2\bm p_T\cdot\bm\ell_T-\ell_T^2+i0}.
 \label{eq:single-contact-main}
\end{align}
Here $f$ contains the Qiu--Sterman function and the remaining factors
that depend on $x_2-x_1$. The soft-gluon poles from both beams select the diagonal
arguments $x_{A2}=x_{A1}$ and $x_{B2}=x_{B1}$, producing
$T_F(x_A,x_A)T_F(x_B,x_B)$. The remaining pole terms are included in
the sum over cuts. We take the real part only after combining the
contributions from both beams, so that the product of their imaginary
parts is included.

When summing the diagrams, we can discard those with vanishing color
factors. For example, if the two coherent
gluons meet at a three-gluon vertex, the color factor of one quark
line is the trace of a single generator, which is zero. Diagrams with
only one crossed gluon attachment also vanish for the same reason.

Since the full finite-$q_T$ result is lengthy, we present only the
leading-power cross section at low $q_T$, which is what enters the
comparison with TMD factorization. To this end, we expand the photon
mass-shell constraint as a distribution:
\begin{equation}
 \delta\left((1-z_A)(1-z_B)-\frac{q_T^2}{\hat{s}}\right)
 \approx\frac{\delta(1-z_B)}{(1-z_A)_+}
 +\frac{\delta(1-z_A)}{(1-z_B)_+}
 +\ln\frac{\hat{s}}{q_T^2}\,
       \delta(1-z_A)\delta(1-z_B).
 \label{eq:longitudinal-distribution}
\end{equation}
The first two terms describe photon radiation collinear to beam $A$ and
beam $B$, respectively. The logarithmic term proportional to
$\delta(1-z_A)\delta(1-z_B)$ arises from the soft region, where the two
collinear regions overlap.

We sum over all photon attachments to the quark lines, all possible
orderings of the coherent-gluon insertions, and all cuts consistent
with the final-state measurement.
The resulting leading-power collinear cross section at low $q_T$ is related
to the reference result by
\begin{equation}
 \frac{d\Delta\Sigma}
 {dy_1\,dy_2\,d^2\bm P_T\,d^2\bm q_T}
 =R\frac{d\Delta\Sigma^{\mathrm{ref}}}
 {dy_1\,dy_2\,d^2\bm P_T\,d^2\bm q_T}.
 \label{eq:combined-leading}
\end{equation}
The sum of $DD$, $TD$, $DT$, and $TT_{\rm same}$ reproduces
Eq.~\eqref{eq:reference-result} in full, including both spin-dependent
tensor structures. The crossed contribution has the same leading-power
kinematic dependence, so it changes only the overall normalization,
not the angular dependence. The resulting $50\%$ enhancement relative
to the SIDIS-based prediction directly quantifies the factorization
breaking caused by color entanglement.

Since Eq.~\eqref{eq:combined-leading} includes the structure of the TMD
prediction up to an overall color factor, we can write down a heuristic
resummation formula by analogy with TMD factorization. The resummed cross section can be cast into the following form
\begin{align}
 &\frac{d\Delta\Sigma_{\mathrm{LL},\gamma}}
 {dy_1\,dy_2\,d^2\bm P_T\,d^2\bm q_T}
 =-\frac{\sigma_0\xi_A\xi_B}{4}
 \left(1+\frac{4}{N_c^2-1}\right)
 \label{eq:joint-photon-resummation}\\
 &\quad\times\int\frac{d^2\bm b_T}{(2\pi)^2}
 e^{i\bm b_T\cdot\bm q_T}
 (\bm b_T\cdot\tilde{\bm S}_A)
 (\bm b_T\cdot\tilde{\bm S}_B)
 \exp\left[-\frac{\alpha_{\mathrm{em}}e_f^2}{2\pi}
 \ln^2\frac{\hat{s}b_T^2}{4e^{-2\gamma_E}}\right]T_F(\xi_A,\xi_A,\mu_b)T_F(\xi_B,\xi_B,\mu_b). \nonumber
\end{align}
Here the scale dependence of the  Qiu--Sterman functions is determined by the QED DGLAP evolution.
Expanding the exponential to first order in $\alpha_{\mathrm{em}}$
reproduces the leading logarithmic term of the fixed-order result in
Eq.~\eqref{eq:combined-leading}.
In this ansatz, the color-blind photon exponent acts on the joint coefficient, not on
two independent Sivers functions.

Extending this ansatz to QCD is less straightforward because soft-gluon
radiation, unlike photon radiation, can mix color structures. Whether
it also generates additional logarithms associated with color
entanglement remains an open question. A step toward restoring
predictive power in the intermediate region would be a resummation
\cite{Collins:1984kg,Idilbi:2004vb,Kang:2011mr,Aybat:2011ge,
Aybat:2011ta,Boer:2013zca,Sun:2013dya,Echevarria:2014xaa}
that incorporates any such logarithms together with the full QCD
evolution of the Qiu--Sterman functions
\cite{Braun:2009mi,Zhou:2015lxa,Ma:2012xn,Kang:2012em,Dai:2014ala,
Henneman:2001ev,Kang:2008ey,Zhou:2008mz,Vogelsang:2009pj,
Ratcliffe:2009pp,Ma:2011nd,Schafer:2012ra,Kang:2012ns,
Schafer:2013opa,Yoshida:2016tfh,Pirnay:2013fra,Rodini:2024usc}.

\section{Summary}
\label{sec:summary}

We have quantified the size of TMD factorization breaking in the
intermediate transverse-momentum region $\Lambda_{\rm QCD}\ll q_T\ll
P_T$, using the double-Sivers asymmetry as an example. The predictive
power of perturbative calculations can thus be partially restored in
this kinematic region.
A key question is whether multiple soft-gluon emissions in QCD generate
additional logarithmic structures associated with color entanglement.
Answering this question and carrying out the corresponding resummation
are natural next steps toward restoring full predictive power in the intermediate $q_T$ region.
For other spin channels, different operator projections and different
coefficients of the pole contributions summed over cuts could modify the
strength of the color-entanglement contribution.
Beyond the perturbative region considered here, it remains unclear
whether the nonperturbative effects of color entanglement at small
$q_T$ admit a controlled quantitative description.

\bigskip
\noindent{\it Acknowledgments.} This work was supported by the National
Natural Science Foundation of China under Grant Nos.~12175118, 12321005,
and 12575091 (J.Z.).

\bibliographystyle{apsrev4-2}
\bibliography{ref}

@article{Rogers:2010dm,
    author = "Rogers, Ted C. and Mulders, Piet J.",
    title = "{No Generalized TMD-Factorization in Hadro-Production of High Transverse Momentum Hadrons}",
    eprint = "1001.2977",
    archivePrefix = "arXiv",
    primaryClass = "hep-ph",
    doi = "10.1103/PhysRevD.81.094006",
    journal = "Phys. Rev. D",
    volume = "81",
    pages = "094006",
    year = "2010"
}

@article{Ji:2006vf,
    author = "Ji, Xiangdong and Qiu, Jian-wei and Vogelsang, Werner and Yuan, Feng",
    title = "{Single Transverse-Spin Asymmetry in Drell-Yan Production at Large and Moderate Transverse Momentum}",
    eprint = "hep-ph/0604023",
    archivePrefix = "arXiv",
    reportNumber = "BNL-NT-06-14, RBRC-574",
    doi = "10.1103/PhysRevD.73.094017",
    journal = "Phys. Rev. D",
    volume = "73",
    pages = "094017",
    year = "2006"
}

@article{Zhou:2009jm,
    author = "Zhou, Jian and Yuan, Feng and Liang, Zuo-Tang",
    title = "{Transverse momentum dependent quark distributions and polarized Drell-Yan processes}",
    eprint = "0909.2238",
    archivePrefix = "arXiv",
    primaryClass = "hep-ph",
    doi = "10.1103/PhysRevD.81.054008",
    journal = "Phys. Rev. D",
    volume = "81",
    pages = "054008",
    year = "2010"
}

@article{Collins:2007nk,
    author = "Collins, John and Qiu, Jian-Wei",
    title = "{$k_{T}$ factorization is violated in production of high-transverse-momentum particles in hadron-hadron collisions}",
    eprint = "0705.2141",
    archivePrefix = "arXiv",
    primaryClass = "hep-ph",
    reportNumber = "ANL-HEP-PR-07-25",
    doi = "10.1103/PhysRevD.75.114014",
    journal = "Phys. Rev. D",
    volume = "75",
    pages = "114014",
    year = "2007"
}

@misc{Collins:2007jp,
    author = "Collins, John",
    eprint = "0708.4410",
    archivePrefix = "arXiv",
    primaryClass = "hep-ph",
    month = "8",
    year = "2007"
}

@article{Rogers:2013zha,
    author = "Rogers, Ted C.",
    title = "{Extra spin asymmetries from the breakdown of transverse-momentum-dependent factorization in hadron-hadron collisions}",
    eprint = "1304.4251",
    archivePrefix = "arXiv",
    primaryClass = "hep-ph",
    doi = "10.1103/PhysRevD.88.014002",
    journal = "Phys. Rev. D",
    volume = "88",
    number = "1",
    pages = "014002",
    year = "2013"
}

@article{Qiu:1998ia,
    author = "Qiu, Jian-wei and Sterman, George F.",
    title = "{Single transverse spin asymmetries in hadronic pion production}",
    eprint = "hep-ph/9806356",
    archivePrefix = "arXiv",
    reportNumber = "ITP-SB-98-28, BNL-HET-98-17",
    doi = "10.1103/PhysRevD.59.014004",
    journal = "Phys. Rev. D",
    volume = "59",
    pages = "014004",
    year = "1999"
}

@article{Boer:2003cm,
    author = "Boer, Daniel and Mulders, P. J. and Pijlman, F.",
    title = "{Universality of T odd effects in single spin and azimuthal asymmetries}",
    eprint = "hep-ph/0303034",
    archivePrefix = "arXiv",
    doi = "10.1016/S0550-3213(03)00527-3",
    journal = "Nucl. Phys. B",
    volume = "667",
    pages = "201--241",
    year = "2003"
}

@article{Bomhof:2004aw,
    author = "Bomhof, C. J. and Mulders, P. J. and Pijlman, F.",
    title = "{Gauge link structure in quark-quark correlators in hard processes}",
    eprint = "hep-ph/0406099",
    archivePrefix = "arXiv",
    doi = "10.1016/j.physletb.2004.06.100",
    journal = "Phys. Lett. B",
    volume = "596",
    pages = "277--286",
    year = "2004"
}

@article{Bomhof:2006dp,
    author = "Bomhof, C. J. and Mulders, P. J. and Pijlman, F.",
    title = "{The Construction of gauge-links in arbitrary hard processes}",
    eprint = "hep-ph/0601171",
    archivePrefix = "arXiv",
    doi = "10.1140/epjc/s2006-02554-2",
    journal = "Eur. Phys. J. C",
    volume = "47",
    pages = "147--162",
    year = "2006"
}

@article{Bomhof:2006ra,
    author = "Bomhof, C. J. and Mulders, P. J.",
    title = "{Gluonic pole cross-sections and single spin asymmetries in hadron-hadron scattering}",
    eprint = "hep-ph/0609206",
    archivePrefix = "arXiv",
    doi = "10.1088/1126-6708/2007/02/029",
    journal = "JHEP",
    volume = "2007",
    number = "02",
    pages = "029",
    year = "2007"
}

@article{Zhou:2009rp,
    author = "Zhou, Jian and Yuan, Feng and Liang, Zuo-Tang",
    title = "{Drell-Yan Lepton Pair Azimuthal Asymmetry in Hadronic Processes}",
    eprint = "0901.3601",
    archivePrefix = "arXiv",
    primaryClass = "hep-ph",
    doi = "10.1016/j.physletb.2009.06.041",
    journal = "Phys. Lett. B",
    volume = "678",
    pages = "264--268",
    year = "2009"
}

@article{Zhou:2010ui,
    author = "Zhou, Jian and Metz, Andreas",
    title = "{Transverse single spin asymmetry in the Drell-Yan process}",
    eprint = "1011.5871",
    archivePrefix = "arXiv",
    primaryClass = "hep-ph",
    doi = "10.1103/PhysRevD.86.014001",
    journal = "Phys. Rev. D",
    volume = "86",
    pages = "014001",
    year = "2012"
}

@article{Ji:2006ub,
    author = "Ji, Xiangdong and Qiu, Jian-Wei and Vogelsang, Werner and Yuan, Feng",
    title = "{A Unified picture for single transverse-spin asymmetries in hard processes}",
    eprint = "hep-ph/0602239",
    archivePrefix = "arXiv",
    reportNumber = "BNL-NT-06-11, RBRC-588",
    doi = "10.1103/PhysRevLett.97.082002",
    journal = "Phys. Rev. Lett.",
    volume = "97",
    pages = "082002",
    year = "2006"
}

@article{Ji:2006br,
    author = "Ji, Xiangdong and Qiu, Jian-Wei and Vogelsang, Werner and Yuan, Feng",
    title = "{Single-transverse spin asymmetry in semi-inclusive deep inelastic scattering}",
    eprint = "hep-ph/0604128",
    archivePrefix = "arXiv",
    reportNumber = "BNL-NT-06-15, RBRC-598",
    doi = "10.1016/j.physletb.2006.05.044",
    journal = "Phys. Lett. B",
    volume = "638",
    pages = "178--186",
    year = "2006"
}

@article{Collins:2002kn,
    author = "Collins, John C.",
    title = "{Leading twist single transverse-spin asymmetries: Drell-Yan and deep inelastic scattering}",
    eprint = "hep-ph/0204004",
    archivePrefix = "arXiv",
    doi = "10.1016/S0370-2693(02)01819-1",
    journal = "Phys. Lett. B",
    volume = "536",
    pages = "43--48",
    year = "2002"
}

@article{Schafer:2014xpa,
    author = {Sch{\"a}fer, Andreas and Zhou, Jian},
    title = "{Color entanglement for {\ensuremath{\gamma}}-jet production in polarized pA collisions}",
    eprint = "1406.3198",
    archivePrefix = "arXiv",
    primaryClass = "hep-ph",
    doi = "10.1103/PhysRevD.90.094012",
    journal = "Phys. Rev. D",
    volume = "90",
    number = "9",
    pages = "094012",
    year = "2014"
}

@article{Zhou:2017mpw,
    author = "Zhou, Jian",
    title = "{Color entanglement like effect in collinear twist-3 factorization}",
    eprint = "1706.02842",
    archivePrefix = "arXiv",
    primaryClass = "hep-ph",
    doi = "10.1103/PhysRevD.96.114001",
    journal = "Phys. Rev. D",
    volume = "96",
    number = "11",
    pages = "114001",
    year = "2017"
}

@article{Boer:2017hqr,
    author = {Boer, Dani{\"e}l and van Daal, Tom and Gaunt, Jonathan R. and Kasemets, Tomas and Mulders, Piet J.},
    title = "{Colour unwound - disentangling colours for azimuthal asymmetries in Drell-Yan scattering}",
    eprint = "1709.04935",
    archivePrefix = "arXiv",
    primaryClass = "hep-ph",
    reportNumber = "NIKHEF-2017-040",
    doi = "10.21468/SciPostPhys.3.6.040",
    journal = "SciPost Phys.",
    volume = "3",
    number = "6",
    pages = "040",
    year = "2017"
}

@article{Qiu:1991pp,
    author = "Qiu, Jian-wei and Sterman, George F.",
    title = "{Single transverse spin asymmetries}",
    reportNumber = "ITP-SB-91-8",
    doi = "10.1103/PhysRevLett.67.2264",
    journal = "Phys. Rev. Lett.",
    volume = "67",
    pages = "2264--2267",
    year = "1991"
}

@article{Qiu:1991wg,
    author = "Qiu, Jian-wei and Sterman, George F.",
    title = "{Single transverse spin asymmetries in direct photon production}",
    reportNumber = "ITP-SB-91-9, ISU-NP-91-21",
    doi = "10.1016/0550-3213(92)90003-T",
    journal = "Nucl. Phys. B",
    volume = "378",
    pages = "52--78",
    year = "1992"
}

@article{Zhou:2015ima,
    author = "Zhou, Jian",
    title = "{Transverse single spin asymmetry in Drell-Yan production in polarized pA collisions}",
    eprint = "1502.02457",
    archivePrefix = "arXiv",
    primaryClass = "hep-ph",
    doi = "10.1103/PhysRevD.92.014034",
    journal = "Phys. Rev. D",
    volume = "92",
    number = "1",
    pages = "014034",
    year = "2015"
}

@article{Zhou:2017sdx,
    author = "Zhou, Jian",
    title = "{Single spin asymmetries in forward p-p/A collisions revisited: the role of color entanglement}",
    eprint = "1704.04901",
    archivePrefix = "arXiv",
    primaryClass = "hep-ph",
    doi = "10.1103/PhysRevD.96.034027",
    journal = "Phys. Rev. D",
    volume = "96",
    number = "3",
    pages = "034027",
    year = "2017"
}

@article{Schafer:2014zea,
    author = {Sch{\"a}fer, Andreas and Zhou, Jian},
    title = "{Transverse single spin asymmetry in direct photon production in polarized pA collisions}",
    eprint = "1404.5809",
    archivePrefix = "arXiv",
    primaryClass = "hep-ph",
    doi = "10.1103/PhysRevD.90.034016",
    journal = "Phys. Rev. D",
    volume = "90",
    number = "3",
    pages = "034016",
    year = "2014"
}

@article{Rein:2025pwu,
    author = {Rein, Daniel and Schlegel, Marc and Tollk{\"u}hn, Patrick and Vogelsang, Werner},
    title = "{Next-to-Leading-Order Corrections and Factorization for Transverse Single-Spin Asymmetries}",
    eprint = "2503.16097",
    archivePrefix = "arXiv",
    primaryClass = "hep-ph",
    doi = "10.1103/fhxq-njl2",
    journal = "Phys. Rev. Lett.",
    volume = "135",
    number = "25",
    pages = "251901",
    year = "2025"
}

@article{Rein:2025qhe,
    author = {Rein, Daniel and Schlegel, Marc and Tollk{\"u}hn, Patrick and Vogelsang, Werner},
    title = "{Transverse nucleon single-spin asymmetry for single-inclusive hadron and jet production at NLO accuracy}",
    eprint = "2503.16119",
    archivePrefix = "arXiv",
    primaryClass = "hep-ph",
    doi = "10.1103/w5cv-ssdm",
    journal = "Phys. Rev. D",
    volume = "112",
    number = "11",
    pages = "114024",
    year = "2025"
}

@article{Scimemi:2019gge,
    author = "Scimemi, Ignazio and Tarasov, Andrey and Vladimirov, Alexey",
    title = "{Collinear matching for Sivers function at next-to-leading order}",
    eprint = "1901.04519",
    archivePrefix = "arXiv",
    primaryClass = "hep-ph",
    doi = "10.1007/JHEP05(2019)125",
    journal = "JHEP",
    volume = "2019",
    number = "05",
    pages = "125",
    year = "2019"
}

@article{Kang:2011mr,
    author = "Kang, Zhong-Bo and Xiao, Bo-Wen and Yuan, Feng",
    title = "{QCD Resummation for Single Spin Asymmetries}",
    eprint = "1106.0266",
    archivePrefix = "arXiv",
    primaryClass = "hep-ph",
    doi = "10.1103/PhysRevLett.107.152002",
    journal = "Phys. Rev. Lett.",
    volume = "107",
    pages = "152002",
    year = "2011"
}

@article{Vogelsang:2009pj,
    author = "Vogelsang, Werner and Yuan, Feng",
    title = "{Next-to-leading Order Calculation of the Single Transverse Spin Asymmetry in the Drell-Yan Process}",
    eprint = "0904.0410",
    archivePrefix = "arXiv",
    primaryClass = "hep-ph",
    doi = "10.1103/PhysRevD.79.094010",
    journal = "Phys. Rev. D",
    volume = "79",
    pages = "094010",
    year = "2009"
}

@article{Henneman:2001ev,
    author = "Henneman, A. A. and Boer, Daniel and Mulders, P. J.",
    title = "{Evolution of transverse momentum dependent distribution and fragmentation functions}",
    eprint = "hep-ph/0104271",
    archivePrefix = "arXiv",
    doi = "10.1016/S0550-3213(01)00557-0",
    journal = "Nucl. Phys. B",
    volume = "620",
    pages = "331--350",
    year = "2002"
}

@article{Kang:2008ey,
    author = "Kang, Zhong-Bo and Qiu, Jian-Wei",
    title = "{Evolution of twist-3 multi-parton correlation functions relevant to single transverse-spin asymmetry}",
    eprint = "0811.3101",
    archivePrefix = "arXiv",
    primaryClass = "hep-ph",
    doi = "10.1103/PhysRevD.79.016003",
    journal = "Phys. Rev. D",
    volume = "79",
    pages = "016003",
    year = "2009"
}

@article{Zhou:2008mz,
    author = "Zhou, Jian and Yuan, Feng and Liang, Zuo-Tang",
    title = "{QCD Evolution of the Transverse Momentum Dependent Correlations}",
    eprint = "0812.4484",
    archivePrefix = "arXiv",
    primaryClass = "hep-ph",
    doi = "10.1103/PhysRevD.79.114022",
    journal = "Phys. Rev. D",
    volume = "79",
    pages = "114022",
    year = "2009"
}

@article{Braun:2009mi,
    author = "Braun, V. M. and Manashov, A. N. and Pirnay, B.",
    title = "{Scale dependence of twist-three contributions to single spin asymmetries}",
    eprint = "0909.3410",
    archivePrefix = "arXiv",
    primaryClass = "hep-ph",
    doi = "10.1103/PhysRevD.80.114002",
    journal = "Phys. Rev. D",
    volume = "80",
    pages = "114002",
    year = "2009",
    note = "[Erratum: Phys.Rev.D 86, 119902 (2012)]"
}

@article{Ratcliffe:2009pp,
    author = "Ratcliffe, Philip G. and Teryaev, Oleg",
    title = "{Non-Universality and Evolution of the Sivers Function}",
    eprint = "0910.5348",
    archivePrefix = "arXiv",
    primaryClass = "hep-ph",
    doi = "10.1142/S0217732309001194",
    journal = "Mod. Phys. Lett. A",
    volume = "24",
    pages = "2984--2994",
    year = "2009"
}

@article{Ma:2011nd,
    author = "Ma, J. P. and Sang, H. Z.",
    title = "{Soft-Gluon-Pole Contribution in Single Transverse-Spin Asymmetries of Drell-Yan Processes}",
    eprint = "1102.2679",
    archivePrefix = "arXiv",
    primaryClass = "hep-ph",
    doi = "10.1007/JHEP04(2011)062",
    journal = "JHEP",
    volume = "2011",
    number = "04",
    pages = "062",
    year = "2011"
}

@article{Schafer:2012ra,
    author = {Sch{\"a}fer, Andreas and Zhou, Jian},
    title = "{A Note on the scale evolution of the ETQS function $T_F(x,x)$}",
    eprint = "1203.5293",
    archivePrefix = "arXiv",
    primaryClass = "hep-ph",
    doi = "10.1103/PhysRevD.85.117501",
    journal = "Phys. Rev. D",
    volume = "85",
    pages = "117501",
    year = "2012"
}

@article{Ma:2012xn,
    author = "Ma, J. P. and Wang, Q.",
    title = "{Scale Dependence of Twist-3 Quark-Gluon Operators for Single Spin Asymmetries}",
    eprint = "1205.0611",
    archivePrefix = "arXiv",
    primaryClass = "hep-ph",
    doi = "10.1016/j.physletb.2012.07.036",
    journal = "Phys. Lett. B",
    volume = "715",
    pages = "157--163",
    year = "2012"
}

@article{Kang:2012em,
    author = "Kang, Zhong-Bo and Qiu, Jian-Wei",
    title = "{QCD evolution of naive-time-reversal-odd parton distribution functions}",
    eprint = "1205.1019",
    archivePrefix = "arXiv",
    primaryClass = "hep-ph",
    doi = "10.1016/j.physletb.2012.06.021",
    journal = "Phys. Lett. B",
    volume = "713",
    pages = "273--276",
    year = "2012"
}

@article{Kang:2012ns,
    author = "Kang, Zhong-Bo and Vitev, Ivan and Xing, Hongxi",
    title = "{Transverse momentum-weighted Sivers asymmetry in semi-inclusive deep inelastic scattering at next-to-leading order}",
    eprint = "1212.1221",
    archivePrefix = "arXiv",
    primaryClass = "hep-ph",
    doi = "10.1103/PhysRevD.87.034024",
    journal = "Phys. Rev. D",
    volume = "87",
    number = "3",
    pages = "034024",
    year = "2013"
}

@article{Dai:2014ala,
    author = "Dai, Ling-Yun and Kang, Zhong-Bo and Prokudin, Alexei and Vitev, Ivan",
    title = "{Next-to-leading order transverse momentum-weighted Sivers asymmetry in semi-inclusive deep inelastic scattering: the role of the three-gluon correlator}",
    eprint = "1409.5851",
    archivePrefix = "arXiv",
    primaryClass = "hep-ph",
    reportNumber = "JLAB-THY-14-1947",
    doi = "10.1103/PhysRevD.92.114024",
    journal = "Phys. Rev. D",
    volume = "92",
    number = "11",
    pages = "114024",
    year = "2015"
}

@article{Yoshida:2016tfh,
    author = "Yoshida, Shinsuke",
    title = "{New pole contribution to $P_{h\perp}$-weighted single-transverse spin asymmetry in semi-inclusive deep inelastic scattering}",
    eprint = "1601.07737",
    archivePrefix = "arXiv",
    primaryClass = "hep-ph",
    doi = "10.1103/PhysRevD.93.054048",
    journal = "Phys. Rev. D",
    volume = "93",
    number = "5",
    pages = "054048",
    year = "2016"
}

@misc{Pirnay:2013fra,
    author = "Pirnay, B. M.",
    title = "{t3evol - Numerical Solution of Twist-three Evolution Equations}",
    eprint = "1307.1272",
    archivePrefix = "arXiv",
    primaryClass = "hep-ph",
    month = "7",
    year = "2013"
}

@article{Rodini:2024usc,
    author = "Rodini, Simone and Rossi, Lorenzo and Vladimirov, Alexey",
    title = "{Numerical implementation of evolution equations for twist-3 collinear PDFs}",
    eprint = "2405.01162",
    archivePrefix = "arXiv",
    primaryClass = "hep-ph",
    reportNumber = "DESY-24-060 ; IPARCOS-UCM-24-025",
    doi = "10.1140/epjc/s10052-024-13060-6",
    journal = "Eur. Phys. J. C",
    volume = "84",
    number = "7",
    pages = "732",
    year = "2024"
}

@article{Idilbi:2004vb,
    author = "Idilbi, Ahmad and Ji, Xiang-dong and Ma, Jian-Ping and Yuan, Feng",
    title = "{Collins-Soper equation for the energy evolution of transverse-momentum and spin dependent parton distributions}",
    eprint = "hep-ph/0406302",
    archivePrefix = "arXiv",
    doi = "10.1103/PhysRevD.70.074021",
    journal = "Phys. Rev. D",
    volume = "70",
    pages = "074021",
    year = "2004"
}

@article{Aybat:2011ge,
    author = "Aybat, S. Mert and Collins, John C. and Qiu, Jian-Wei and Rogers, Ted C.",
    title = "{The QCD Evolution of the Sivers Function}",
    eprint = "1110.6428",
    archivePrefix = "arXiv",
    primaryClass = "hep-ph",
    reportNumber = "NIKHEF-2011-030, YITP-SB-11-37",
    doi = "10.1103/PhysRevD.85.034043",
    journal = "Phys. Rev. D",
    volume = "85",
    pages = "034043",
    year = "2012"
}

@article{Aybat:2011ta,
    author = "Aybat, S. Mert and Prokudin, Alexei and Rogers, Ted C.",
    title = "{Calculation of TMD Evolution for Transverse Single Spin Asymmetry Measurements}",
    eprint = "1112.4423",
    archivePrefix = "arXiv",
    primaryClass = "hep-ph",
    reportNumber = "JLAB-THY-11-1466, NIKHEF-2011-033, YITP-SB-11-48",
    doi = "10.1103/PhysRevLett.108.242003",
    journal = "Phys. Rev. Lett.",
    volume = "108",
    pages = "242003",
    year = "2012"
}

@article{Boer:2013zca,
    author = "Boer, Daniel",
    title = "{TMD evolution of the Sivers asymmetry}",
    eprint = "1304.5387",
    archivePrefix = "arXiv",
    primaryClass = "hep-ph",
    doi = "10.1016/j.nuclphysb.2013.05.021",
    journal = "Nucl. Phys. B",
    volume = "874",
    pages = "217--229",
    year = "2013"
}

@article{Sun:2013dya,
    author = "Sun, Peng and Yuan, Feng",
    title = "{Energy Evolution for the Sivers Asymmetries in Hard Processes}",
    eprint = "1304.5037",
    archivePrefix = "arXiv",
    primaryClass = "hep-ph",
    doi = "10.1103/PhysRevD.88.034016",
    journal = "Phys. Rev. D",
    volume = "88",
    pages = "034016",
    year = "2013"
}

@article{Echevarria:2014xaa,
    author = "Echevarria, Miguel G. and Idilbi, Ahmad and Kang, Zhong-Bo and Vitev, Ivan",
    title = "{QCD Evolution of the Sivers Asymmetry}",
    eprint = "1401.5078",
    archivePrefix = "arXiv",
    primaryClass = "hep-ph",
    reportNumber = "NIKHEF-2014-004",
    doi = "10.1103/PhysRevD.89.074013",
    journal = "Phys. Rev. D",
    volume = "89",
    pages = "074013",
    year = "2014"
}

@article{Ma:2014uma,
    author = "Ma, J. P. and Zhang, G. P.",
    title = "{Gauge Invariance and QCD Twist-3 Factorization for Single Spin Asymmetries}",
    eprint = "1409.2938",
    archivePrefix = "arXiv",
    primaryClass = "hep-ph",
    doi = "10.1007/JHEP02(2015)163",
    journal = "JHEP",
    volume = "2015",
    number = "02",
    pages = "163",
    year = "2015"
}

@article{Chen:2016dnp,
    author = "Chen, A. P. and Ma, J. P. and Zhang, G. P.",
    title = "{One-loop corrections to single spin asymmetries at twist-3 in Drell-Yan processes}",
    eprint = "1607.08676",
    archivePrefix = "arXiv",
    primaryClass = "hep-ph",
    doi = "10.1103/PhysRevD.95.074005",
    journal = "Phys. Rev. D",
    volume = "95",
    number = "7",
    pages = "074005",
    year = "2017"
}

@article{Chen:2017lvx,
    author = "Chen, A. P. and Ma, J. P. and Zhang, G. P.",
    title = "{One-Loop Corrections of Single Spin Asymmetries in Semi-Inclusive DIS}",
    eprint = "1708.09091",
    archivePrefix = "arXiv",
    primaryClass = "hep-ph",
    doi = "10.1103/PhysRevD.97.054003",
    journal = "Phys. Rev. D",
    volume = "97",
    number = "5",
    pages = "054003",
    year = "2018"
}

@article{Qiu:2007ar,
    author = "Qiu, Jian-Wei and Vogelsang, Werner and Yuan, Feng",
    title = "{Asymmetric di-jet production in polarized hadronic collisions}",
    eprint = "0704.1153",
    archivePrefix = "arXiv",
    primaryClass = "hep-ph",
    reportNumber = "BNL-NT-07-15, RBRC-665",
    doi = "10.1016/j.physletb.2007.05.023",
    journal = "Phys. Lett. B",
    volume = "650",
    pages = "373--378",
    year = "2007"
}

@article{Qiu:2007ey,
    author = "Qiu, Jian-Wei and Vogelsang, Werner and Yuan, Feng",
    title = "{Single Transverse-Spin Asymmetry in Hadronic Dijet Production}",
    eprint = "0706.1196",
    archivePrefix = "arXiv",
    primaryClass = "hep-ph",
    reportNumber = "BNL-NT-07-25, RBRC-676",
    doi = "10.1103/PhysRevD.76.074029",
    journal = "Phys. Rev. D",
    volume = "76",
    pages = "074029",
    year = "2007"
}

@article{Collins:1984kg,
    author = "Collins, John C. and Soper, Davison E. and Sterman, George F.",
    title = "{Transverse Momentum Distribution in Drell-Yan Pair and W and Z Boson Production}",
    reportNumber = "CERN-TH-3923",
    doi = "10.1016/0550-3213(85)90479-1",
    journal = "Nucl. Phys. B",
    volume = "250",
    pages = "199--224",
    year = "1985"
}

@article{Schafer:2013wca,
    author = {Sch{\"a}fer, Andreas and Zhou, Jian},
    title = "{Transverse single spin asymmetry in hadronic {\ensuremath{\eta}}c,b production}",
    eprint = "1302.4600",
    archivePrefix = "arXiv",
    primaryClass = "hep-ph",
    doi = "10.1103/PhysRevD.88.014008",
    journal = "Phys. Rev. D",
    volume = "88",
    number = "1",
    pages = "014008",
    year = "2013"
}

@article{Zhou:2015lxa,
    author = "Zhou, Jian",
    title = "{Note on the scale dependence of the Burkardt sum rule}",
    eprint = "1507.02819",
    archivePrefix = "arXiv",
    primaryClass = "hep-ph",
    doi = "10.1103/PhysRevD.92.074016",
    journal = "Phys. Rev. D",
    volume = "92",
    number = "7",
    pages = "074016",
    year = "2015"
}

@misc{Schafer:2013opa,
    author = {Sch{\"a}fer, Andreas and Zhou, Jian},
    title = "{A note on the scale evolution of tri-gluon correlations}",
    eprint = "1308.4961",
    archivePrefix = "arXiv",
    primaryClass = "hep-ph",
    month = "8",
    year = "2013"
}

@article{Brodsky:2002rv,
    author = "Brodsky, Stanley J. and Hwang, Dae Sung and Schmidt, Ivan",
    title = "{Initial state interactions and single spin asymmetries in Drell-Yan processes}",
    eprint = "hep-ph/0206259",
    archivePrefix = "arXiv",
    reportNumber = "SLAC-PUB-9271, USM-TH-126",
    doi = "10.1016/S0550-3213(02)00617-X",
    journal = "Nucl. Phys. B",
    volume = "642",
    pages = "344--356",
    year = "2002"
}

\end{document}